\documentclass{aa}
\usepackage{epsf}

\begin{document}

\title{Thermal structure and cooling of neutron stars \\
         with  magnetized envelopes}
         
 \titlerunning{Thermal structure and cooling of magnetic neutron stars}

\author{A. Y. Potekhin
\and
 D. G. Yakovlev}

\offprints{A. Y. Potekhin\\
\email{palex@astro.ioffe.rssi.ru}}

\institute{Ioffe Physico-Technical Institute,
         Politekhnicheskaya 26, 194021 St.~Petersburg, Russia}

\date{Received 19 February 2001 / Accepted 17 April 2001}
\abstract{
The thermal structure of neutron stars with magnetized envelopes
is studied using modern physics input.
The relation between the internal ($T_\mathrm{int}$)
and local surface temperatures is calculated
and fitted by analytic expressions
for magnetic field strengths $B$ from 0 to $10^{16}$~G
and arbitrary inclination of the field lines to the surface.
The luminosity of a neutron star with dipole magnetic field
is calculated and fitted as a function
of $B$, $T_\mathrm{int}$, stellar mass and radius.
In addition, we simulate cooling
of neutron stars with magnetized envelopes.
In particular, we analyse ultramagnetized envelopes
of magnetars and also the effects of the magnetic field
of the Vela pulsar on the determination of
critical temperatures of neutron and proton superfluids
in its core.
}
\maketitle

\section{Introduction}
\label{sect-intro}
It is well known that theoretical models of
cooling of isolated neutron stars (NSs) depend on the
poorly known equation of state (EOS) of
superdense matter in the NS
interiors. Comparing calculated cooling curves
(decrease of the NS effective surface
temperature $T_\mathrm{e}$ in time) with observations
gives a potentially powerful method of testing microscopic theories
of superdense matter
(e.g., Pethick \cite{pethick}; Page \cite{page97,page98}).
Its practical implementation is restricted by
the accuracy of observational determination of $T_\mathrm{e}$
and NS ages and by the quality of the physics input
used in calculations. Great observational
progress has been achieved recently
after the launch of \emph{Chandra} and \emph{Newton} X-ray observatories.
In the present paper we update
theoretical models of thermal structure
and evolution of NSs with magnetized envelopes.

Most of the observed NSs 
possess magnetic fields $B\sim10^{12}$--$10^{13}$~G
(e.g., Taylor et al.\ \cite{tml93}).
Some NSs are possibly magnetars, with $B>10^{14}$~G
(e.g., Thompson \& Duncan \cite{td95};
Kouveliotou et al.\ \cite{kou-ea98,kou-ea99};
Mereghetti \cite{mereghetti01}).
The internal NS magnetic field can be even higher.
The strong magnetic field affects physical
properties of all NS layers in many ways.
For instance, the field $B \sim 10^{16}$ G
may affect the neutrino emissivity in the NS core
(e.g., Baiko \& Yakovlev \cite{by99}). The field $B \ga 10^{12}$ G
may change the electrical resistivity of NS cores,
accelerating evolution of the internal field. In principle
the thermal evolution may be coupled to the
magnetic one (e.g., Urpin \& Shalybkov \cite{us95}).

Thus modelling of the thermal structure
and evolution of the magnetized NSs is a complicated
task. In this paper we focus on two problems.
First we consider the thermal structure of the
outer magnetized NS envelope of density
$\rho < \rho_\mathrm{b} = 4 \times 10^{11}$ g cm$^{-3}$
with the magnetic field $B \la 10^{16}$ G.
These envelopes produce thermal insulation
(blanketing) of NS interiors.
We solve this problem using updated physics
input described in Sect.\ \ref{sect-input}.
The solution (Sect.\ \ref{sect-thst}) relates
the internal NS temperature $T_\mathrm{int}$
with the local effective surface temperature $T_\mathrm{s}$
and, therefore, with the integrated NS luminosity (or
the mean effective surface temperature, $T_\mathrm{e}$). 
Second, in order to illustrate
the obtained $T_\mathrm{s}(T_\mathrm{int})$ relation,
we simulate (Sect.\ \ref{sect-cool})
cooling of NSs which possess a
given dipole or radial magnetic field 
in the outer envelopes.

The effects of a strong magnetic field
on thermodynamic and kinetic properties
of the outer NS layers
have been reviewed, for instance, by
Yakovlev \& Kaminker (\cite{YaK}) and
Ventura \& Potekhin (\cite{elounda}).
The field affects the properties
of all plasma components, the electron
component usually being affected most strongly.
Motion of free electrons perpendicular to the field lines
is quantized in Landau orbitals
with a characteristic transverse scale equal to
the \emph{magnetic length}
$a_\mathrm{m}=(\hbar c/eB)^{1/2}=\lambda_\mathrm{e}/\sqrt{b}$,
where $\lambda_\mathrm{e}=\hbar/(m_\mathrm{e} c)$
is the electron Compton wavelength,
$b=\hbar\omega_\mathrm{c}/m_\mathrm{e} c^2 = B_{12}/44.14$ 
is the magnetic field strength  expressed in relativistic units,
$\omega_\mathrm{c}={eB/m_\mathrm{e} c}$
is the electron cyclotron frequency,
and $B_{12}\equiv B/(10^{12}$~G).
Except for the outermost parts of the NS envelopes,
the electrons constitute degenerate, almost ideal gas.
The electron entropy,
magnetization, thermal and electrical conductivities,
and other quantities
exhibit quantum oscillations of the de Haas--van Alphen type.
The oscillations occur under variations of density $\rho$ or $B$
whenever the electron Fermi momentum reaches the characteristic values
$\sqrt{2nb}\,m_\mathrm{e} c$
associated with occupation of Landau levels
$n=1,2,\ldots$
These oscillations appreciably change the properties
of degenerate electrons in the
limit of a {\it strongly quantizing\/} field
(e.g., Yakovlev \& Kaminker \cite{YaK})
in which almost all electrons populate the ground Landau level.
The latter case takes place 
at temperature $T\ll T_B$ and density $\rho<\rho_B$, where
\begin{eqnarray}
   \rho_B & = & m_\mathrm{u} n_B \, A/Z
  \approx 7045
       \,B_{12}^{3/2}\,(A/Z)\mbox{~g\,cm$^{-3}$},
\label{rho_B}
\\
    T_B & = &
    {\hbar\omega_\mathrm{g}/k_\mathrm{B}} \approx
             {1.343\times10^8\, (B_{12}/\sqrt{1+x_\mathrm{r}^2})}{\rm~K}.
\label{T_B}
\end{eqnarray}
Here, $m_\mathrm{u} = 1.66054\times10^{-24}$~g is the atomic mass unit,
$A$ and $Z$ 
are the mean ion mass and charge numbers,
$\omega_\mathrm{g}=\omega_\mathrm{c}/\sqrt{1+x_\mathrm{r}^2}$ 
is the electron gyrofrequency,
\begin{equation}
   x_\mathrm{r} = {{\hbar(3\pi^2n_\mathrm{e})^{1/3}}\over{m_\mathrm{e} c}}
   \approx
   1.009 \,(\rho_6\,Z/ A )^{1/3}
   \label{x_r}
\end{equation}
is the \emph{non-magnetic
relativity parameter}, $\rho_6\equiv\rho/(10^6\mbox{~g\,cm$^{-3}$})$,
and 
$n_B=1/(\pi^2\sqrt2\,a_\mathrm{m}^3)$ is the electron number density
at which the Fermi energy reaches the first excited Landau level.
In the case of \emph{a weakly quantizing magnetic field}
($\rho \ga \rho_B$, $T \la T_B$), the oscillations of
electron quantities occur
around the classical non-magnetic values and are typically
not very pronounced.
In the case of $T \gg T_B$,
the field can be treated as classical (\emph{non-quantizing})
and the oscillations disappear.

The thermal structure of magnetized NS
envelopes has been studied
by a number of authors mainly using a
plane-parallel approximation.
More attention has been paid to the
case of the radial magnetic field (normal to the surface).
It has been thoroughly considered by Hernquist (\cite{hern85}),
Van Riper (\cite{kvr88}), Schaaf (\cite{schaaf90a}),
Heyl \& Hernquist (\cite{hh-multi}) for $B \la 10^{14}$ G.
In another paper
Heyl \& Hernquist (\cite{hh-theory})
analysed the case of higher fields, $B \sim 10^{15}$--$10^{16}$ G.
One can consult the cited papers for references to
earlier works. The principal conclusion
of these studies is that
the magnetic field reduces thermal insulation
of the blanketing envelope by increasing
the longitudinal (along the field lines) thermal
conductivity of degenerate electrons due to Landau
quantization of electron motion.

The thermal structure of the envelope with the magnetic field
tangential to the surface has been analysed by
Hernquist (\cite{hern85}), Schaaf (\cite{schaaf90a}),
and Heyl \& Hernquist (\cite{hh-theory}) for $B \la 10^{14}$ G.
In this case the field increases thermal insulation
of the blanketing envelope due to the classical
effect of reduction of the electron thermal conductivity
transverse to the field because of the Larmor rotation.

The case of arbitrary inclination of the field to
the surface was studied by Greenstein \&
Hartke (\cite{gh83}) in the approximation
of constant (density and temperature independent)
longitudinal and transverse thermal conductivities.
The authors proposed a very simple formula (Sect.\ \ref{sect-Tvar})
which relates the local  
 surface and internal stellar temperatures, $T_\mathrm{s}$
and $T_\mathrm{int}$.
It is constructed from two $T_\mathrm{s}(T_\mathrm{int})$ 
relations obtained
for the radial and tangential magnetic fields.
Page (\cite{page95}) presented arguments that the formula
of Greenstein \& Hartke is valid also for realistic,
variable thermal conductivities. If so, one
immediately gets the required $T_\mathrm{s}(T_\mathrm{int})$
relation for any magnetic field
inclination using the realistic relations for the radial
and tangential fields.
This method has been used by several authors
(e.g., Shibanov \& Yakovlev \cite{shibyak}).
Recently, the case of arbitrary field inclination
has been studied also by Heyl \& Hernquist (\cite{hh-multi}).
In Sect.\ \ref{sect-thst} we reconsider the
thermal structure of the blanketing envelopes
for any magnetic field inclination and compare our results
with those of earlier studies.

Early simulations of cooling of magnetized NSs
were performed assuming the radial magnetic field
everywhere over the stellar surface
(e.g., Nomoto \& Tsuruta \cite{nt87}; Van Riper \cite{kvr91}).
Since the radial magnetic field reduces the thermal insulation,
these theories predicted acceleration of cooling
of the magnetized NS accompanied by enhanced
NS luminosity at the early (neutrino-dominated) cooling stage
(at stellar ages $t \la 10^4 - 10^5$~yr).
Page (\cite{page95}) and
Shibanov \& Yakovlev (\cite{shibyak}) simulated cooling
of NSs with dipole magnetic fields.
They showed that the decrease of thermal
insulation of the stellar envelope near the magnetic
equator was partly compensated by its increase near the pole,
and the magnetic field did not necessarily accelerate the cooling,
in agreement with an earlier conjecture of Hernquist (\cite{hern85}).
In a series of papers Heyl \& Hernquist
(\cite{HH97,HH97b,hh-theory,hh-multi}) proposed simplified
models of cooling of magnetized NSs including the cases
of ultrahigh surface magnetic fields, $B \sim 10^{15}-10^{16}$ G.

We illustrate our new models of magnetized NS envelopes
with simulation of cooling of NSs (Sect.\ \ref{sect-cool}).
We briefly discuss
cooling of ultramagnetized NSs as well as cooling models
of the Vela pulsar with the dipole magnetic field and
superfluid core.

\section{Physics input}
\label{sect-input}
\subsection{Equations of thermal structure and evolution}
The internal hydrostatic structure of a NS can be regarded as
temperature-independent (e.g., Shapiro \& Teukolsky \cite{ST83}).  
It is conventional (Gudmundsson et al.\ \cite{gpe83}) to separate the
calculation of heat transport in the NS interior
(at radius $r < R_\mathrm{b}$)
and in the outer heat-blanketing envelope
($R_\mathrm{b} \leq r \leq R$, where
$R$ is the stellar radius). The choice of the boundary
$R_\mathrm{b}$ must meet several requirements.
The blanketing envelope should be thin
($R-R_\mathrm{b} \ll R$) and contain negligibly small mass;
there should be no large sources of energy generation or sink 
present there;
it should serve as a good thermal blanket of
the internal region; its thermal relaxation time should
be shorter than time-scales of temperature variation
in the internal region. 
In the studies of non-magnetized NSs,
the boundary is usually taken at 
the density $\rho_\mathrm{b}=10^{10}$ g cm$^{-3}$
(e.g., Gudmundsson et al.\ \cite{gpe83}; Potekhin et al.\ \cite{pcy97}
-- hereafter PCY).
In the presence of a strong magnetic field
we assume
additionally that
the temperature
does not vary over the boundary $r=R_\mathrm{b}$
at any given moment.
To make this requirement more realistic,
we shift the boundary to the neutron drip density,
$\rho_\mathrm{b}=4\times10^{11}\mbox{~g\,cm$^{-3}$}$.

\subsubsection{Internal region}
In the internal region ($r<R_\mathrm{b}$), 
magnetic fields $B \la 10^{16}$~G are either non-quantizing or
weakly quantizing (degenerate electrons populate many Landau levels).
For simplicity, we neglect the effects
of magnetic fields in this region and
use a spherically symmetric temperature
distribution which obeys the classical
equations of thermal evolution (Thorne \cite{thorne}):
\begin{equation}
    { \mathrm{e}^{-2\Phi } \over 4 \pi r^2 } \,
    \sqrt{1 - {2 G m  \over c^2 r}} \,
    { \partial  \over \partial  r}
    \left( \mathrm{e}^{2 {\Phi }} L \right)
    = -Q_\nu - {C_v  \mathrm{e}^{-\Phi }} \,
     {\partial T \over \partial t},
  \label{cool-therm-balance}
\end{equation}
\begin{equation}
    {L \over 4 \pi r^2} =
    -  \kappa \,\sqrt{1 - {2Gm  \over c^2 r}} \, \mathrm{e}^{-{\Phi }}
    {\partial \over \partial r} \left( T \mathrm{e}^{\Phi } \right),
  \label{cool-Fourier}
\end{equation}
where $Q_\nu(r)$ is the neutrino emissivity, 
$C_v(r)$ is the heat capacity, $\kappa(r)$ is the
thermal conductivity, $L(r)$ is the local luminosity defined as
the non-neutrino heat flux transported through a sphere of radius $r$.
Furthermore, $G$ is the gravitational constant,
$m(r)$ is the gravitational mass inside the sphere of radius $r$,
and $\Phi(r)$ is the metric function
determined by the stellar model.  
Notice that at the stellar surface
$\mathrm{e}^{\Phi(R)}=(1-r_g/R)^{-1/2}$,
where
$r_g=2GM/c^2=2.95(M/M_\odot)$ km is the Schwarzschild radius
defined by the total gravitational NS mass $m(R)=M$.
On the right-hand side of 
Eq.~(\ref{cool-therm-balance}) we neglect
the rate of heat production (e.g., by internal friction
due to differential rotation, see Page \cite{page97,page98}
and references therein). This
effect is important for relatively
old and cold NSs, which we will not consider here.
To solve Eqs.~(\ref{cool-therm-balance}) and (\ref{cool-Fourier}),
we use the computer code described 
by Gnedin et al.\ (\cite{gyp}, hereafter GYP).
After thermal relaxation within the NS,
at $t \ga 100$ yr the
redshifted temperature 
$\widetilde{T}(t) \equiv T(r,t)\mathrm{e}^{\Phi(r)}$ 
becomes constant throughout the internal region.

\subsubsection{Blanketing envelope}
\label{sect-blank}
The thermal structure of the blanketing envelope is studied
in the stationary, local plane-parallel approximation to relate the
local effective surface temperature $T_\mathrm{s}$
to the temperature $T_\mathrm{int}$ at the
inner boundary of the envelope. Everywhere on the surface,
except possibly in tiny regions
where the field is almost tangential,
one may assume that a scale of
temperature variation over the surface is much larger than the
thickness of the blanketing envelope.
This leads to the \emph{one-dimensional} approximation
for the heat diffusion equation:
\begin{equation}
 F \equiv \sigma T_\mathrm{s}^4 = \kappa {{\rm\,d}T\over{\rm\,d}z}=
      {16\sigma\over3}{T^3{\rm\,d}T\over{\rm\,d}\tau},
\quad
    \kappa\equiv{16\sigma T^3\over3K\rho},
\label{Flx}
\end{equation}      
where $F$ is the local heat flux density (constant throughout
a given local part of the envelope),
$\kappa$ is an \emph{effective} thermal conductivity
along the normal to the surface, $\sigma$ is
the Stefan--Boltzmann constant, 
$K$ is the mean opacity (Sect.\ 2.3),
       $\tau=\int_{-\infty}^z K\rho \, \,\mathrm{d} z$ is the
optical depth,
and
$z=(R-r)\,\mathrm{e}^{\Phi(R)}$ is the local proper depth.

Integration of Eq.~(\ref{Flx}) gives a temperature profile
$
 T \approx T_\mathrm{s} (\frac34\tau+\frac12)^{1/4},
$
and the integration constant corresponds to the Eddington 
approximation ($\tau=\frac23$ at the {\em radiative surface\/},
where $T=T_\mathrm{s}$).
A more accurate boundary condition 
requires solution of the radiative transfer equation in 
the NS atmosphere (e.g., Shibanov et al.\ \cite{shibetal98}).

At $z\ll R$
the general relativistic equation of hydrostatic equilibrium
can be reduced to the Newtonian form: $\mathrm{d} P/\mathrm{d} z=g\rho$,
where $g=GM/(R^2 \sqrt{1-r_g/R}\,)$ is the surface gravity.
Together with Eq.~(\ref{Flx}), it leads to the thermal structure equation
\begin{equation}
   {\mathrm{d}\log T\over\mathrm{d}\log P} =
   {3\over 16}\,{PK\over g}\,{T_\mathrm{s}^4\over T^4},
\label{th-str}
\end{equation}
where $P$ is the pressure.

The thermal conductivity tensor of magnetized plasma is anisotropic.
It is characterized by the conductivities parallel ($\kappa_\|$)
and transverse ($\kappa_\perp$) to the field, and by the 
off-diagonal (Hall) component.
In the plane-parallel approximation Eq.~(\ref{Flx})
contains the effective thermal conductivity
\begin{equation}
   \kappa=\kappa_\|\cos^2\theta + \kappa_\perp\sin^2\theta,
\end{equation}
where $\theta$ is the angle between the field lines 
and the normal to the surface.
In case $\theta=0$ the heat is transported solely
by the parallel conductivity, $\kappa=\kappa_\|$,
while in case $\theta=90^\circ$ it is transported
by the transverse conductivity, $\kappa=\kappa_\perp$.
These special cases will be referred to as the \emph{parallel} ($\|$)
and \emph{transverse} ($\perp$) conduction cases. 
Actually, $B$ and $\theta$ vary slowly over the surface.
The total stellar luminosity is
\begin{equation}
   L= 
 \int F \,\mathrm{d}\Sigma =
\sigma \int T_\mathrm{s}^4 \,\mathrm{d}\Sigma
   \equiv 4 \pi \sigma R^2 T_\mathrm{e}^4,
\label{L}
\end{equation}
where $\mathrm{d}\Sigma=R^2 \mathrm{d}\Omega$ is the surface
element determined by corresponding solid angle $\mathrm{d}\Omega$,
and $T_\mathrm{e}$ is the (mean) effective 
temperature to be distinguished from the local effective
temperature $T_\mathrm{s}$. 
Naturally, $T_\mathrm{e}=T_\mathrm{s}$ for
a non-magnetic NS.

Assuming the dipole field, we
can use the general-relativistic solution
(Ginzburg \& Ozernoy \cite{GinzOz})
\begin{eqnarray}
&&
  B(\chi) = B_\mathrm{p}\,\sqrt{\cos^2\chi+a^2\sin^2\chi},
  \label{dipole}
\quad
  \mathrm{tan}\,\theta = a\,\mathrm{tan}\,\chi,
\\&&
  a = - {(1-x) \ln(1-x) + x - 0.5 x^2
        \over [ \ln(1-x)+x + 0.5 x^2] \sqrt{1-x}},
\end{eqnarray}
where $B_\mathrm{p}$ is the field strength at the magnetic pole,
$\chi$ is the polar angle, and $x=r_g/R$.

The quantities $T_\mathrm{s}$, $T_\mathrm{e}$ and $L$
refer to a local reference frame at the NS surface.
The redshifted (``apparent'') quantities
as detected by a distant observer are (Thorne \cite{thorne}):
$T_\mathrm{s}^\infty = T_\mathrm{s} \, \sqrt{1-r_g/R}$,
$T_\mathrm{e}^\infty = T_\mathrm{e} \, \sqrt{1-r_g/R}$,
$L_\infty = (1-r_g/R)\, L$.

Crucial for the thermal evolution is the relation between
$T_\mathrm{s}$ and $T_\mathrm{int}=T(R_\mathrm{b})$. 
For non-magnetic blanketing envelopes composed of iron,
this relation was studied by Gudmundsson et al. (\cite{gpe83}),
while the non-magnetic envelopes of various chemical compositions
were analysed by PCY.
Ventura \& Potekhin (\cite{elounda})
presented an analytic description of
the $T_\mathrm{int}(T_\mathrm{s})$ relation
for the envelopes without magnetic fields and with
strong magnetic fields; Heyl \& Hernquist (\cite{hh-theory})
performed a 
semianalytic investigation of the case of strong magnetic field.
The results of Ventura \& Potekhin (\cite{elounda})
and Heyl \& Hernquist (\cite{hh-theory})
are rather approximate because of a number
of simplifying assumptions discussed
by Ventura \& Potekhin (\cite{elounda}).

The validity of the one-dimensional approximation can be checked
by direct two-dimensional simulation of the heat transport 
in the blanketing envelope. 
Such simulation has been attempted by Schaaf (\cite{schaaf})
for a homogeneously magnetized NS
under many simplified assumptions, so that
a more realistic study is required.
The heat conduction
from hotter to cooler zones along the surface 
or possible meridional and convective motions
can smooth the temperature variations over the NS surface.
Nevertheless, the one-dimensional approximation seems to be
sufficient for our cooling calculations presented below.

\subsection{Equation of state}
In the inner region of a NS,
the effects of magnetic field are assumed to be
weak and we use
the same EOSs as in GYP.
Specifically, in the core, we use
the moderately stiff phenomenological EOS of
matter composed of neutrons, protons and electrons, as proposed by
Prakash et al.\ (\cite{pal88}), in its simplified version suggested
by Page \& Applegate (\cite{pa92}).
In the inner envelope, we use the model of
ground-state matter (Negele \& Vautherin \cite{nv73})
and describe the properties of atomic nuclei
by the smooth composition model 
(Kaminker et al.\ \cite{kaminker99}). The
core--crust interface is placed at $\rho=1.5 \times 10^{14}$ g
cm$^{-3}$.  

The outer envelope is assumed to be composed of iron,
which can be partially ionized at $\rho\la10^6\mbox{~g\,cm$^{-3}$}$.
Following PCY, 
we employ the mean-ion approximation
and adjust an effective ion charge 
to a more elaborate EOS:
the Opacity Library (OPAL) 
EOS (Rogers et al.\ \cite{OPAL_EOS}) at $B=0$ and
the Thomas--Fermi EOS of Thorolfsson et al.\ (\cite{Thorolf})
at $B=(10^{10}$--$10^{13}$)~G.
The tabular entries of these EOSs in $\rho$ and $T$ are interpolated
in the same way as in PCY.
When necessary, we interpolate the effective charge at intermediate $B$.
Since no reliable EOS of iron has been published for $B>10^{13}$~G,
we use the effective charge obtained at $B=10^{13}$~G for
higher $B$.

Thus, in our model of the outer envelope, the pressure
is produced by magnetized Fermi gas of electrons and
by the gas of classical ions
with an effective charge.
We neglect the anomalous magnetic moments
of the nuclei, which is a good approximation at $B\ll10^{16}$~G
(cf.\ Broderick et al.\ \cite{bpl00}; Suh \& Mathews \cite{suh}).
The electron pressure can be expressed through
the chemical potential $\mu$ and $T$ 
(e.g., Blandford \& Hernquist \cite{bh82}): 
\begin{equation}
   P_\mathrm{e} =
   P_r\,{b \tau_0^{3/2}\over\sqrt2\pi^2}
   \sum_{n=0}^{n_\mathrm{max}} (2-\delta_{n0})
           (1+2bn)^{1/4} I_{1/2}(\chi_n,\tau_n),
\label{presmag}
\end{equation}
where 
\[
   \chi_n = {\mu-m_\mathrm{e} c^2\,\sqrt{1+2bn} \over k_B T},
\qquad
   \tau_n = {1\over\sqrt{1+2bn}}\,{T\over T_\mathrm{r}},
\]
$P_\mathrm{r}=m_\mathrm{e} c^2/\lambda_\mathrm{e}^3
\approx1.4218\times10^{25}$ dyn cm$^{-2}$
and $T_\mathrm{r}=m_\mathrm{e} c^2/k_\mathrm{B}\approx5.930 \times10^9$~K
are the relativistic units of pressure and temperature, respectively.
The standard Fermi--Dirac integral
$  I_{1/2} $
is given by a fit presented in Chabrier \& Potekhin (\cite{cp98}).
The chemical potential $\mu$ is found at given $\rho$, $T$, and $B$
using an analytic fit and iterative procedure described by 
Potekhin \& Yakovlev (\cite{py96}).

\subsection{Opacities}
\label{sect-opa}
The heat is carried through the NS envelope
mainly by electrons at relatively high densities
and by photons near the surface (e.g., Gudmundsson et al.\ \cite{gpe83}).
In general, the two mechanisms work in parallel, hence
\begin{equation}
   \kappa=\kappa_\mathrm{r}+\kappa_\mathrm{c},
\qquad
 K^{-1} = {K}_\mathrm{r}^{-1} + {K}_\mathrm{c}^{-1}, 
 \label{op}
\end{equation}      
where $\kappa_\mathrm{r}$, $\kappa_\mathrm{c}$ 
and ${K}_\mathrm{r}$, ${K}_\mathrm{c}$ denote
the radiative (r) and electron (c)
components of the conductivity and opacity. 

Typically, the radiative conduction dominates
($\kappa_\mathrm{r}>\kappa_\mathrm{c}$) in 
the outermost non-degenerate layers of a NS,
whereas the electron conduction 
dominates ($\kappa_\mathrm{c}>\kappa_\mathrm{r}$)
in deeper, moderately and strongly degenerate layers.
The $T_\mathrm{s}(T_\mathrm{int})$ 
relation depends mainly on the conductivities in the
\emph{sensitivity strip} on the $\rho-T$ plane
(Gudmundsson et al.\ \cite{gpe83})
placed near the \emph{turning point},
where $\kappa_\mathrm{c}\sim\kappa_\mathrm{r}$.

\subsubsection{Radiative opacities}
Radiative opacities of partially ionized iron in strong magnetic fields
were studied by Rajagopal et al.\ (\cite{RRM}), but 
neither tables nor analytic fits required
for NS modelling were published.
Fortunately, we will see that
the $T_\mathrm{s}(T_\mathrm{int})$
relation for a not too cold NS
is not noticeably changed if we replace
the radiative opacity of partially ionized iron by that of 
fully ionized iron.
Therefore we base our opacity model on the assumption of 
full ionization. In this case, the two principal contributions
into the opacity
come from free-free transitions 
in electron-ion collisions and from Thomson
scattering of photons by free electrons.

At $B=0$, the second process is described by
the well-known Thomson cross section $\sigma_\mathrm{T}$
in the non-relativistic limit $T\ll T_\mathrm{r}$
(e.g., Berestetski\u{\i} et al.\ \cite{QED}).
The corresponding opacity is 
\begin{equation}
   K_\mathrm{T}={n_\mathrm{e}\sigma_\mathrm{T} \over \rho} =
         {8\pi\over3}\,\left({e^2\over m_\mathrm{e} c^2}\right)^2
         {n_\mathrm{e} \over \rho}.
 \label{Thomson}
\end{equation}

The free-free absorption coefficient
at $B=0$ was calculated
by Karzas \& Latter (\cite{KarzasLatter})
as a function of the electron
velocity and photon frequency $\omega$.
Hummer (\cite{Hummer88}) produced an accurate fit to
its thermal average in the range $10^{-4}\leq \xi \leq 10^{1.5}$ 
and $10^{-3}\leq T_\mathrm{Ry}<10^3$,
where 
\begin{equation}
   T_\mathrm{Ry}\equiv {2\hbar^2\over m_\mathrm{e} Z^2 e^4}\,k_\mathrm{B} T = 
     {T_6\over0.157\,89\,Z^2},
\quad
   \xi \equiv{\hbar\omega\over k_\mathrm{B} T}.
\label{TRy}
\end{equation}
The ratio of the 
Rosseland mean, $K_\mathrm{ff}$, calculated using that fit,
to the Thomson opacity can be written as
\begin{equation}
   {K_\mathrm{ff}\over K_\mathrm{T}} =
    {Z\alpha_\mathrm{f}\over3c_7}\left({2\over\pi}\right)^{3/2}
        \left({T_\mathrm{r}\over T}\right)^{7/2} x_\mathrm{r}^3
      \approx{2\times10^4\over c_7}\,{Z^2\over A}\,
          {\rho\over T_6^{7/2}},
\label{KRcompar}
\end{equation}
where $\rho$ is measured in g cm$^{-3}$, 
$\alpha_\mathrm{f}=e^2/(\hbar c)\approx\frac{1}{137}$,
\begin{equation}
   c_7 = {15\over4\pi^4}\int_0^\infty{\xi^7\,e^{-\xi}\over
        (1-e^{-\xi})^3\,\Lambda_\mathrm{ff}(T,\xi)}\,\mathrm{d} \xi,
\label{c7}
\end{equation}
$\Lambda_\mathrm{ff}=(\pi/\sqrt{3})\,\bar{g}_\mathrm{ff}$ is the Coulomb
logarithm, and $\bar{g}_\mathrm{ff}$ is the thermally
averaged Gaunt factor.
In the Born approximation
(e.g., Silant'ev \& Yakovlev \cite{SilYak}),
$\Lambda_\mathrm{ff}=e^{\xi/2}K_0(\xi/2)$, where $K_0$ is a modified
Bessel function; then $c_7=316.824$. Since 
$e^{\xi/2}K_0(\xi/2) \approx \sqrt{\pi/\xi}$ at $\xi\to\infty$,
we can obtain a reasonable approximation to $\bar{g}_\mathrm{ff}$
by extending the
accurate fit of Hummer (\cite{Hummer88}) beyond its boundary
$\xi=10^{1.5}$ in the power law form
$\bar{g}_\mathrm{ff}\propto \xi^{-1/2}$.
Then we can calculate the integral in Eq.~(\ref{c7}).
The result is fitted (within 0.6\%) by
\begin{equation}
   c_7={108.8+77.6\,T_\mathrm{Ry}^{0.834}
    \over 1+0.502\,T_\mathrm{Ry}^{0.355}+0.245\,T_\mathrm{Ry}^{0.834}}.
\end{equation}

If free-free transitions and scattering at $B=0$ are 
important simultaneously,
the Rosseland mean of the sum of the two opacities
can be presented as
\begin{equation}
   K_\mathrm{r}=(K_\mathrm{ff}+K_\mathrm{T})\,A(f,T),
\quad
   f\equiv K_\mathrm{ff}/(K_\mathrm{ff}+K_\mathrm{T}).
\label{KR}
\end{equation}
The non-additivity factor $A(f,T)$ has
been calculated by Silant'ev \& Yakovlev (\cite{SilYak})
in the Born approximation
(in which case it does not depend on $T$).
Beyond the Born approximation, we obtain a function weakly
depending on $T$,
which can be fitted (within 1\%) as
\begin{equation}
   A(f,T) = 1+{1.097+0.777\,T_\mathrm{Ry} \over 1+0.536\,T_\mathrm{Ry}}
           \,f^{0.617}\,(1-f)^{0.77}.
\label{Afit}
\end{equation}

\begin{figure}
\centering
\epsfxsize=86mm
\epsffile[115 240 445 660]{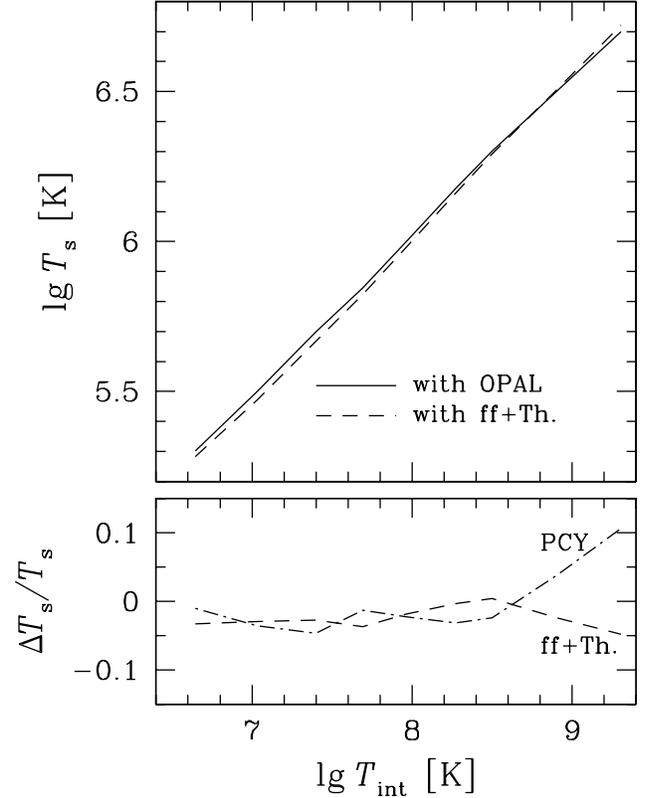}
\caption{$T_\mathrm{s}(T_\mathrm{int})$
relations at $B=0$ and $g=2.43 \times 10^{14}$ cm s$^{-2}$, 
obtained using OPAL radiative opacities
(solid line) and the free-free+Thomson
opacities described in the text (dashed line).
Lower window:
fractional differences between the accurate numerical 
values of $T_\mathrm{s}(T_\mathrm{int})$
and those
obtained using (i) the free-free+Thomson opacity
(dashed line) and (ii)
the older analytic formula of PCY (dot-dashed line).
}
\label{fig-tbts_chk}
\end{figure}

Thus we can propose a simplified analytic treatment
of the radiative opacity at $B=0$ as the opacity
produced by free-free transitions and Thomson scattering alone.
The effect of this simplified treatment
is illustrated in Fig.~\ref{fig-tbts_chk}.
Here, we have applied the code for solving 
the thermal structure equation, described in PCY, with 
the updated conductive opacities (Potekhin et al.\ \cite{pbhy99})
and two different sets of radiative opacities.
The solid line on the upper panel
is obtained using the OPAL radiative opacities
(Iglesias \& Rogers \cite{OPAL}), whereas the dashed line
is obtained using Eqs.\ (\ref{Thomson})--(\ref{Afit}).
The resulting difference (lower panel) 
does not exceed several percent
in the most important range of $T_\mathrm{int}$ shown in the figure.
The difference between the curve obtained
using Eqs.\ (\ref{Thomson})--(\ref{Afit})
and the fit
derived in PCY (dot-dashed line) is of similar magnitude.
The latter difference arises not only from the fit error (PCY),
but also from
the improvement of the electron conduction opacity 
(Potekhin et al.\ \cite{pbhy99}) and, at $T_\mathrm{int}\ga10^9$~K,
from shifting $\rho_\mathrm{b}$
to $4\times10^{11}\mbox{~g\,cm$^{-3}$}$ (for, as noted 
in PCY, isothermality
is not fully reached at $\rho=10^{10}\mbox{~g\,cm$^{-3}$}$ for high $T$).

Interaction of photons with magnetized plasma
depends on their polarization and propagation direction.
Accordingly, radiative thermal conductivities
and associated opacities along and across the field
become different.
Rosseland mean opacities
$K_{\mathrm{r}\|}$ and $K_{\mathrm{r}\perp}$
were calculated by Silant'ev \& Yakovlev (\cite{SilYak})
for various values
of $T$ and $B$.
We have fitted their numerical results by
\begin{eqnarray}
   { K_\mathrm{r}(\rho,T,0)\over K_{\mathrm{r}\|}(\rho,T,B)} &=& 1+
      {A_1\,u+(A_2\,u)^2 \over 1+A_3\,u^2}\,u^2,
\label{KRmag-l}
\\
   { K_\mathrm{r}(\rho,T,0)\over K_{\mathrm{r}\perp}(\rho,T,B)} &=& {1+
      (A_4\,u)^{3.5}+(A_5\,u)^4 \over 1+A_6\,u^2},
\label{KRmag-t}
\\
  A_n &=& a_n - b_n\,f^{\,c_n},
\label{KRmag}
\end{eqnarray}
where $u=T_B/(2T)$,
$K_\mathrm{r}(\rho,T,0)$ is the field-free opacity (\ref{KR}),
and the parameter $f$ introduced in Eq.~(\ref{KR})
is calculated at $B=0$.
The best-fit parameters $a_n$, $b_n$ and $c_n$
given in Table~\ref{tab-KRmag} ensure an average
fit error of 5.5\% with the maximum error of 11\%.

\begin{table}
\caption{Coefficients $a_n$, $b_n$ and $c_n$
of Eq.~(\protect\ref{KRmag}).
}
\label{tab-KRmag}
\begin{center}
\begin{tabular}{lllllll}
\hline
$n$ & 1 & 2 & 3 & 4 & 5 & 6\\
\hline
$a_n$ & 0.0949 & 0.1619 & 0.2587 & 0.3418 & 0.4760 & 0.2533\\
$b_n$ & 0.0610 & 0.1400 & 0.1941 & 0.0415 & 0.3115 & 0.1547\\
$c_n$ & 0.090 & 0.0993 & 0.0533 & 2.15 & 0.2377 & 0.231 \\
\hline
\end{tabular}
\end{center}
\end{table}

Asymptotic behaviour of Eqs.~(\ref{KRmag-l}) and (\ref{KRmag-t})
at $u\to\infty$ agrees with theoretical results of Silant'ev \& Yakovlev
(\cite{SilYak}): 
$K_{\mathrm{r}\|}(\rho,T,B)=2K_{\mathrm{r}\perp}(\rho,T,B)
                        =(\pi/u)^2\,K_\mathrm{r}(\rho,T,0)$
at $f=0$ (Thomson scattering);
$K_{\mathrm{r}\|}(\rho,T,B)=K_{\mathrm{r}\perp}(\rho,T,B)$
at $f=1$ (free-free transitions);
and  $K_\mathrm{r}\propto u^{-2}$ at any $f$.

At finite but large $u$
the radiative opacities of fully ionized matter 
are strongly reduced.
The reduction is $\sim10$ times
stronger for scattering than for free-free transitions.
Inserting this factor 10 into Eq.~(\ref{KRcompar})
and taking into account that,
at $B\ga10^{11}$~G, the NS radiative surface
is pushed to $\rho_\mathrm{s} \sim B_{12}\mbox{~g\,cm$^{-3}$}$
(Ventura \& Potekhin \cite{elounda}),
one can see that in deep, strongly magnetized photospheric layers
Thomson scattering dominates the opacity only
at $T_6\ga10\,\rho^{2/7} \ga 10\,B_{12}^{2/7}$.

We note that the Rosseland
mean opacity due to \emph{scattering by free ions}, 
$K_\mathrm{r}^{\rm(s,i)}$,
can be obtained from that for electrons,  $K_\mathrm{T}$,
by the simple scaling:
\begin{equation}
   K_\mathrm{r}^{\rm(s,i)}(\rho,T,B) \approx Z^3
          \left({m_\mathrm{e}\over A m_\mathrm{u}}\right)^2
          \,K_\mathrm{T}\left(\rho,T,{Zm_\mathrm{e}\over A m_\mathrm{u}}\,B\right).
\label{Kion}
\end{equation}
On the contrary, there is no simple scaling
for the opacity of photons due to electron-ion
(electron bremsstrahlung) and ion-ion (ion bremsstrahlung) collisions.
For instance, the processes are well known to
be drastically different in the electric dipole approximation.

{}From the asymptote of $K_\mathrm{T}$ at large $u$ we see that
the Thomson opacity of ions
is $\sim Z$ times larger than that of electrons,
if the ions are strongly quantized into Landau levels,
which occurs at $T_6 \ll 0.07\,(Z/A)\,B_{12}$.
Since the main contribution into the radiative opacity
at sufficiently low temperature 
comes from free-free processes (see above),
we can conclude that 
the Thomson ion scattering cannot contribute appreciably
to the Rosseland mean opacity of the photosphere
{\em unless\/} $B_{12}\gg10^3\,(A/Z)^{1.4}$.

\subsubsection{Electron conductivities}
\label{sect-el-cond}
Electron thermal and electrical
conduction is the most important process
that determines thermal structure  
and magnetic evolution of NSs.
In dense, strongly coupled Coulomb plasmas, 
typical of NS envelopes, 
the electron thermal conductivity is mainly determined by
electron scattering off ions. At $B=0$
it can be written as
\begin{equation}
   \kappa  =  { \pi^2 k_\mathrm{B}^2 T n_\mathrm{e}  \tau_\kappa
                 \over 3 m_\mathrm{e}^\ast},
\label{dg-kin-coeff}
\end{equation} 
where $m_\mathrm{e}^\ast \equiv m_\mathrm{e} \, \sqrt{1+x_\mathrm{r}^2}$
is the kinematic electron mass
and $\tau_\kappa$ is the effective relaxation time.

Recently, Baiko et al.\ (\cite{baiko-ea98})
considerably improved the
treatment of $\kappa$ by taking into account
multiphonon absorption and emission processes
in Coulomb crystals and
incipient quasiordering in a Coulomb liquid of ions.
Using these results, Potekhin et al.\ (\cite{pbhy99})
constructed an effective scattering potential
which allowed them to calculate the non-magnetic
conductivity in the relaxation time approximation
in an analytic form; these analytic expressions
described accurately numerical results obtained beyond
the framework of the relaxation time approximation.
The effective potential has been
used then at arbitrary magnetic fields in order to derive
practical expressions for evaluation of
electrical and thermal conductivities
of degenerate electrons
in magnetized outer envelopes of NSs 
(Potekhin \cite{P99}).
Unlike previous treatments of the electrical conductivities
perpendicular to the quantizing magnetic fields
(Kaminker \& Yakovlev \cite{kayak}; Hernquist \cite{hern84};
Schaaf \cite{schaaf88}), Potekhin (\cite{P99})
went beyond the assumption that $\omega_\mathrm{g}\tau_\kappa\gg1$
by introducing an interpolation from the case
 $\omega_\mathrm{g}\tau_\kappa \gg 1$ 
to $\omega_\mathrm{g}\tau_\kappa \ll 1$.
All tensor components of the kinetic coefficients in magnetic fields
have been calculated and fitted by analytic formulae; 
the corresponding Fortran
code is available electronically at
http://www.ioffe.rssi.ru/astro/conduct/.
This code is used here to calculate the temperature profile
in the outer envelope of a NS.

In the \emph{inner} envelope, we need to know 
the thermal conductivity $\kappa$ only in young NSs,
$t \la 100$ yr, as long as the internal region ($r<R_\mathrm{b}$)
remains non-isothermal (GYP).
For this purpose, we will use
the non-magnetic 
conductivities presented by GYP.
They generalize the expressions
by Potekhin et al.\ (\cite{pbhy99}) in two respects.
First, the atomic nuclei in the inner envelope
cannot be considered as pointlike:
the size and shape of nuclear charge distribution
significantly affect the conductivity.
Second, the electron-phonon 
scattering in the inner crust at temperatures much below $10^8$~K
changes its character: the so called {\em umklapp\/} processes cease to
dominate and the {\em normal\/} processes (with electron momentum
transfer within one Brillouin zone) become more important.

\begin{figure}
\centering
\epsfxsize=8cm
\epsffile[80 275 550 590]{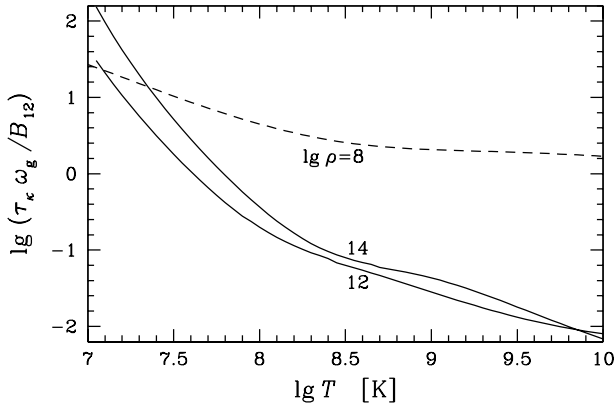}
\caption{Electron magnetization parameter
at the top and bottom of the inner crust
(solid lines) and in the outer envelope (dashed line).
The curves are marked with the values of $\lg \rho$
[g cm$^{-3}$].
}
\label{fig-relaxt}
\end{figure}

The internal temperature in young NSs
is relatively high, $T \ga 10^8$~K.
Our neglect of the effects of the magnetic fields
in the internal region
is strictly justified for the case of small
magnetization parameter, $\omega_\mathrm{g} \tau_\kappa<1$.
This parameter
is shown in Fig.~\ref{fig-relaxt} as a function of $T$
for three values of density: near the top 
and the bottom of the inner envelope (solid lines) and in
the middle of the outer envelope (dashed line).
We see that the use of non-magnetic $\kappa$
is justified for the NSs with the
magnetic fields $B\la10^{13}$~G in the inner envelope.
However it cannot be justified
for higher $B$ in the inner envelope or
for $B \ga 10^{10}$ G in the outer envelope.
In such cases the conductivity
across the field is suppressed by the large factor
$[1+(\omega_\mathrm{g} \tau_\kappa)^2]$ and
the Hall thermal conductivity may become important.
We neglect these effects in the inner envelope.
Therefore, our simulations
(Sect.\ 4) of cooling of young NSs ($t \la 100$ yr)
are justified for $B \la 10^{13}$ G
in the inner envelopes. The results for older NSs,
for which the inner envelope is isothermal,
are valid at larger $B$ as well.

\section{Thermal structure}
\label{sect-thst}
%

\subsection{Temperature profiles}
\label{sect-profiles}
We have solved Eq.~(\ref{th-str}) using the numerical algorithm
described in PCY with the EOS and opacities
presented in the previous section. 
We have varied the magnetic field
inclination angle $\theta$ from 0 to $90^\circ$,
the effective temperature $T_\mathrm{s}$ from $2\times10^5$~K
to $10^7$~K,
and the field strength $B$ from 0 to $10^{16}$~G.
As clear from the discussion in Sect.\ \ref{sect-input},
our model EOS and opacities
may be crude at $B \ga 10^{14}$~G.
Further improvements of physics input are required
for obtaining more reliable results at such $B$.

\begin{figure}
\centering
\epsfxsize=\hsize
 \epsffile[45 330 340 515]{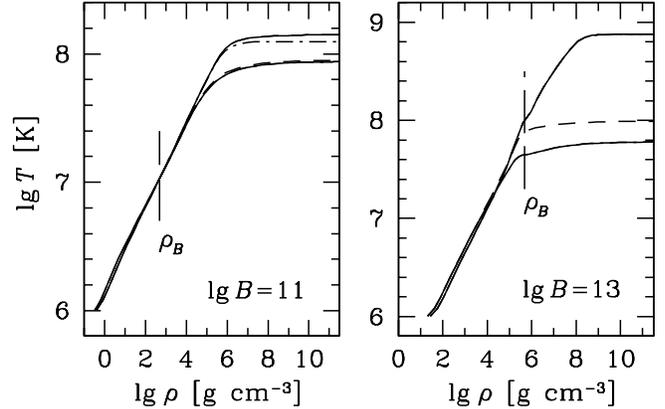}
\caption{Temperature profiles through an iron envelope of
a NS with $M=1.4\,M_\odot$ and $R=10$ km
for the magnetic field $B=10^{11}$~G (left panel)
and $10^{13}$~G (right panel) at
the fixed effective surface temperature $T_\mathrm{s}=10^6$~K.
Lower solid curves correspond to
$\theta=0$ and upper ones to $\theta=90^\circ$.
Dashed lines are calculated with the classical thermal conductivity
for $\theta=0$,
while dot-dashed ones are calculated for $\theta=90^\circ$ assuming
$\omega_\mathrm{g}\tau_\kappa\gg1$ which is
traditional for the transverse conduction.
 }
\label{fig-profcomp}
\end{figure}

\begin{figure*}
\centering
\epsfxsize=17cm
 \epsffile[45 330 565 500]{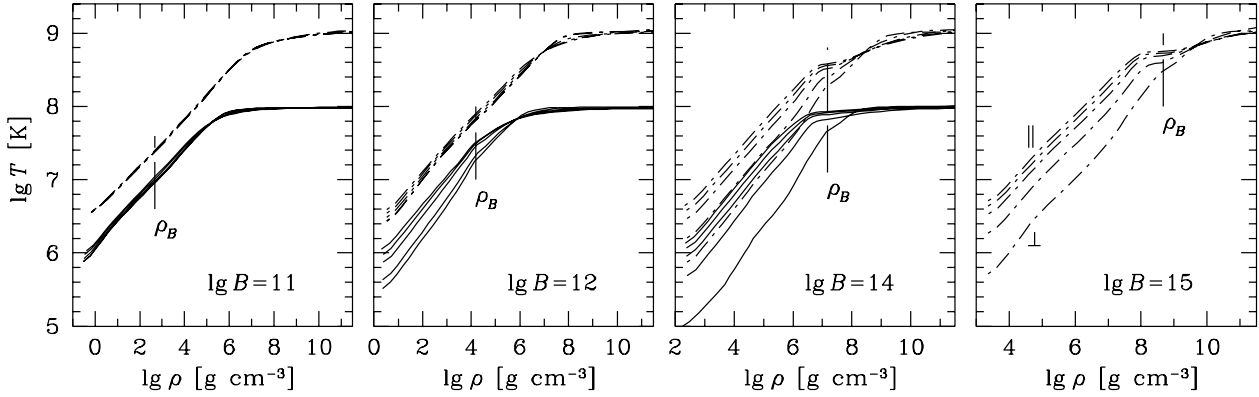}
\caption{Temperature profiles through an iron envelope of
a NS with $M=1.4\,M_\odot$ and $R=10$ km
at (from left to right)
$B=10^{11}$, $10^{12}$, $10^{14}$,
and $10^{15}$~G.
The internal temperature is fixed to $T_\mathrm{int}=10^8$~K
(solid lines) or $10^9$~K (dot-dashed lines).
The lines of each group correspond to
$\cos\theta=1$ (the lowest line),
0.7, 0.4, 0.1, and 0 (the highest line).
 }
\label{fig-profiles}
\end{figure*}

Figures~\ref{fig-profcomp} and \ref{fig-profiles}
present the calculated
$T(\rho)$ profiles for a NS of mass $M=1.4M_\odot$
and radius $R=10$ km 
($g=2.43\times10^{14}{\rm~cm~s}^{-2}$
and $r_g/R=0.413$). The curves in Fig.\ \ref{fig-profcomp}
are calculated at fixed $T_\mathrm{s}=10^6$~K.
The solid curves show the results of accurate calculations
for the magnetic field normal and tangential to the surface
(cases of parallel and transverse conduction, Sect.\ \ref{sect-blank}).
The dashed lines are obtained for the parallel conduction using
the classical thermal conductivity
(neglecting the Landau quantization).
This approximation becomes inaccurate with increasing $B$.
The dot-and-dash lines are calculated
for the tangential magnetic field
using the transverse conductivity in the limit of
$\omega_\mathrm{g}\tau_\kappa\gg1$, employed in previous papers
(e.g., Hernquist \cite{hern84,hern85}; 
Schaaf \cite{schaaf88,schaaf90a,schaaf};
Heyl \& Hernquist \cite{hh-theory,hh-multi}).
This approximation is inaccurate at lower $B$;
a more accurate approximation (solid line) is given by 
the interpolation (Potekhin \cite{P99}) 
mentioned in Sect.~\ref{sect-el-cond}.

Figure \ref{fig-profiles} shows temperature profiles
at four field strengths $B$ and five inclinations $\theta$
for $T_\mathrm{int}=10^8$ and $10^9$~K
(we do not present the curves for $T_\mathrm{int}=10^8$
at $B=10^{15}$ G
because the temperature $T(\rho)$ is
too small near the surface in the transverse conduction case).
The curves show pronounced $\theta$-dependence
at high densities starting from
the turning point (Sect.~\ref{sect-opa}).

The endpoints of the curves in Figs.~\ref{fig-profcomp}
and \ref{fig-profiles}
lie at the \emph{radiative surface} $\rho=\rho_\mathrm{s}$,
where $T=T_\mathrm{s}$, and near the neutron drip point.
Their behaviour qualitatively agrees 
with the results of the approximate analytic study of
Ventura \& Potekhin (\cite{elounda}).
For instance, we confirm
the approximate linear dependence $\rho_\mathrm{s}\propto B$
and the shift of the turning point to higher densities
with increasing $B$.
The density $\rho_B$ marked on the graphs
is defined by Eq.\ (\ref{rho_B}). It
separates the strongly- and weakly-quantizing field regimes.
We see that the increase
of $T$ at $\rho > \rho_B$
(neglected by Heyl \& Hernquist \cite{hh-theory})
can be significant.

The higher the temperature, the wider
is the density region,
where $\kappa_\|$ and $\kappa_\perp$ are of similar magnitude.
Therefore the dependence of the profiles on the inclination $\theta$
and magnetic field strength $B$
is less pronounced at higher $T_\mathrm{int}$
showing convergence to the $B=0$ case.

\subsection{Relation between internal and effective temperatures}
A simple analytic fit to $T_\mathrm{s}(T_\mathrm{int})$
was found by Gudmundsson et al.\ (\cite{gpe83})
for iron envelopes at $T_\mathrm{s} > 3 \times 10^5$~K.
Later, using an updated physics input, PCY constructed
a more general fit for NSs with (accreted) envelopes
containing layers of different chemical elements. In all cases,
the scaling law $T_\mathrm{s} \propto g^{1/4}$ has been confirmed.
This approximate law,
first noted by Gudmundsson et al.\ (\cite{gpe83}),
follows from the thermal structure equation (\ref{th-str}),
if one takes into account that the largest contribution to
the integral of this equation comes from the
sensitivity strip, 
being almost independent of the outer boundary
condition (cf.\ Ventura \& Potekhin \cite{elounda}).
The scaling $T_\mathrm{s} \propto g^{1/4}$
holds also in our case of magnetized envelopes.
Therefore the ratio $\,\mathcal{X}$ of the surface temperature
$T_\mathrm{s}(B,\theta,g,T_\mathrm{int})$ at a given magnetic
field to the value $T_\mathrm{s}^{(0)}(g,T_\mathrm{int})$
in the absence
of the field is practically independent of $g$:
\begin{equation}
   T_\mathrm{s}(B,\theta,g,T_\mathrm{int})\approx
   T_\mathrm{s}^{(0)}(g,T_\mathrm{int}) \,\mathcal{X}(B,\theta,T_\mathrm{int}).
 \label{X}
\end{equation}
Since our outer envelope consists of iron,
$T_\mathrm{s}^{(0)}$ is given by Eq.\ (A7) of PCY:
\begin{equation}
   T_\mathrm{s}^{(0)} \approx  10^6 \,
   g_{14}^{1/4}\left[(7\zeta)^{2.25}+(\zeta/3)^{1.25}\right]^{1/4}~~{\rm K},
\label{PCY-iron}
\end{equation}
where
$\zeta\equiv T_\mathrm{int,9}
-0.001\,g_{14}^{1/4}\,\sqrt{7 \, T_\mathrm{int,9}}$,
$T_\mathrm{int,9}\equiv T_\mathrm{int}/(10^9$~K)
and $g_{14}\equiv g/10^{14}{\rm~cm~s}^{-2}$.

For the parallel and transverse conduction
cases (at the magnetic pole and equator)
we have, respectively,
\begin{eqnarray}
  \mathcal{X}_\|(B,T_\mathrm{int}) &\approx&
           1+0.0492\,B_{12}^{0.292}/T_\mathrm{int,9}^{0.240},
\label{fit1}
\\
  \mathcal{X}_\perp(B,T_\mathrm{int}) &\approx&
          {\sqrt{1+0.1076\,B_{12} (0.03+T_\mathrm{int,9})^{-0.559}}
       \over
             [1+0.819\,B_{12} / (0.03+T_\mathrm{int,9})]^{0.6463}}.
\label{fit2}
\end{eqnarray}
These fits have been checked for $B \la 10^{16}$~G
and $10^7 \mbox{ K} \leq T_\mathrm{int} \leq 10^{9.5}$~K
with an additional constraint
$T_\mathrm{s}>2\times10^5$~K.
Maximum residuals (11\% and 5\%, respectively)
occur at the largest $B$.

\begin{figure*}
\centering
\sidecaption
\epsfxsize=12cm
 \epsffile[30 250 565 535]{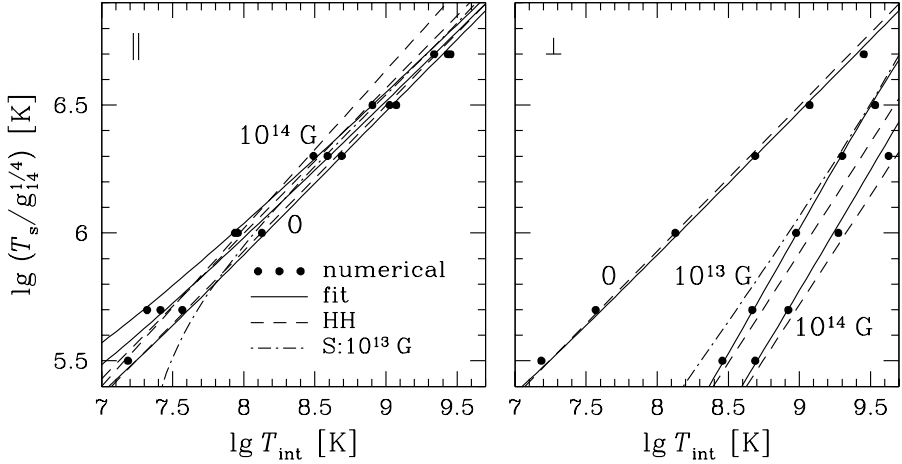}
\caption{~Dependence of the surface temperature $T_\mathrm{s}$ 
on the internal temperature $T_\mathrm{int}$
at the magnetic fields $B=0$, $10^{13}$, and $10^{14}$~G
normal (left panel) or tangential (right panel)
to the surface. Numerical data (heavy dots)
are compared with the present
fit (\protect\ref{X})--(\protect\ref{fit2})
(solid lines) 
and with the earlier fits 
of Schaaf (\protect\cite{schaaf90a}) (dot-dashed line,
for $B=10^{13}$~G)
and Heyl \& Hernquist
(\protect\cite{hh-multi}) (dashed lines).
}
\label{fig-tt}
\end{figure*}
 
The effect of the magnetic field on the effective
temperature and the quality of the present fit
are illustrated in Fig.~\ref{fig-tt}.
The magnetic field normal to the surface enhances the
photon luminosity, 
since heat is transported along the field
lines ($\|$).
Accordingly, $T_\mathrm{s}$ grows up with increasing $B$.
In contrast, the tangential magnetic field
significantly decreases $T_\mathrm{s}$ since the heat propagates across
the field lines ($\perp$).
Our results are in qualitative agreement with earlier results
of Schaaf (\cite{schaaf90a}) and Heyl \& Hernquist
(\cite{hh-multi}), also shown in Fig.\ \ref{fig-tt},
although there are quantitative differences.
The origin of the large difference from Schaaf (\cite{schaaf90a})
is not clear, whereas
the difference from the results of Heyl \& Hernquist
(\cite{hh-multi}) can be attributed to our more accurate
treatment of electron thermal conductivity,
particularly in the nondegenerate region of the envelope.
Note that the results of Heyl \& Hernquist (\cite{hh-multi})
for parallel conduction do not
converge to the $B=0$ results in the limit of high $T_\mathrm{int}$. 
In addition to the above authors,
the case of parallel conduction was
considered thoroughly by Van Riper (\cite{kvr88}) who however
did not present any tables or fit expressions 
for practical applications. We have compared our results
with those presented in his Fig.\ 29. In a hot NS
($T_\mathrm{int} \ga 3 \times 10^9$ K) his results are
in good agreement with ours and satisfy the criterion of
convergence to the $B=0$ case. For $B= 0$ the agreement is
also quite good at any $T_\mathrm{int} \ga 10^7$ K
($T_\mathrm{s} \ga 3 \times 10^5$ K)
but for $B \ga 10^{13}$ G and
$T_\mathrm{int} \ll 3 \times 10^9$ K the magnetized
envelopes of Van Riper appear to be much more heat 
transparent than ours (in contrast, the envelopes
of Heyl \& Hernquist \cite{hh-multi} are less
transparent than ours at the same conditions).
The nature of this difference is also unclear.
 
\subsection{Variation of temperature over stellar surface}
\label{sect-Tvar}
The dependence of $T_\mathrm{s}$ on the angle
$\theta$ between the magnetic field and the normal
to the surface is most easily
described by the model of Greenstein \& Hartke (\cite{gh83}),
which implies a superposition of ``longitudinal"
and ``transverse" heat fluxes:
$T_\mathrm{s}^4(\theta)=T_\mathrm{s}^4(0)\cos^2\theta 
+T_\mathrm{s}^4(90^\circ)\sin^2\theta $.
This approximation has been used, 
e.g., by Page (\cite{page95}); Shibanov \& Yakovlev (\cite{shibyak});
Heyl \& Hernquist (\cite{hh-theory}).
Our numerical calculations confirm that it reproduces accurately
(within $\approx30$\%) the $T_\mathrm{s}(\theta)$ dependence.
However, a replacement of the power index 4 with 4.5
yields better accuracy (within 10\%):
\begin{eqnarray}
 \mathcal{X}(B,\theta,T_\mathrm{int}) & = &
      \big[ \,\mathcal{X}_\|^{9/2}(B,T_\mathrm{int})  \cos^2\theta
\nonumber\\&&
   + \,\mathcal{X}_\perp^{9/2}(B,T_\mathrm{int}) \sin^2\theta \big]^{2/9}.
 \label{fit3}
\end{eqnarray}
Still better accuracy can be reached by varying the power index
between 4 and 5 as a function of $B$ and $T_\mathrm{int}$.

\begin{figure*}
\centering
\sidecaption
\epsfxsize=12cm
 \epsffile[70 265 565 550]{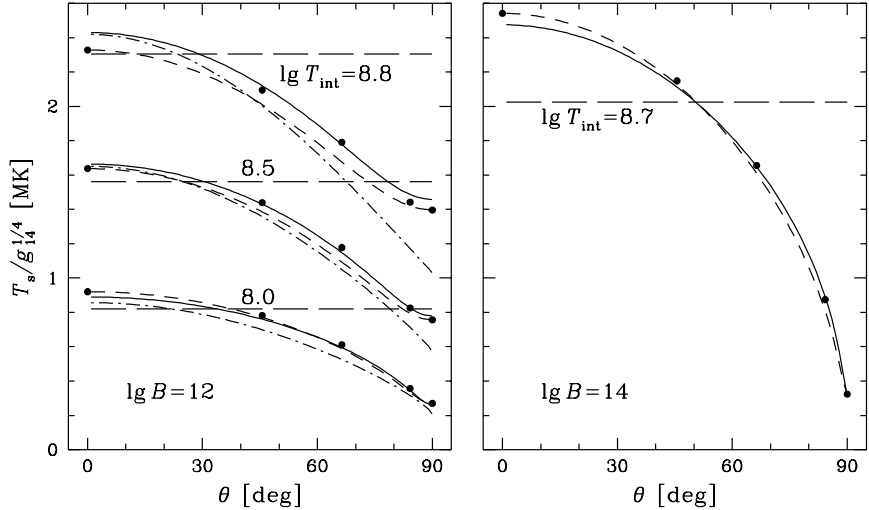}
\caption{~Dependence of the surface temperature $T_\mathrm{s}$ 
on the inclination angle $\theta$ at constant $B$ and $T_\mathrm{int}$.
Left panel: $B=10^{12}$~G, $\lg\,T_\mathrm{int}\mbox{ [K]}=8.0$, 8.5, 8.8;
right panel: $B=10^{14}$~G, $\lg\,T_\mathrm{int}\mbox{ [K]}=8.7$.
Dots: numerical results; solid lines: fit;
dashed lines: interpolation of Greenstein--Hartke
between the accurate values at $\theta=0$ and $90^\circ$;
long dashes: the case of $B=0$; dot-dashed lines on the left panel:
combined fitting formulae of Schaaf (1990a,1990b).
}
\label{fig-tdistr}
\end{figure*}

Figure \ref{fig-tdistr} shows the distribution of the effective
temperature from the magnetic pole to the equator. 
Heavy dots show the results of numerical integration
of $T(\rho)$ profiles.
Solid lines are obtained using our fit,
Eqs.~(\ref{X})--(\ref{fit3}),
while short-dashed lines correspond to the
approximation of Greenstein--Hartke (\cite{gh83}).
For comparison, dot-dashed lines
on the left panel represent the fitting formulae 
of Schaaf (\cite{schaaf90a,schaaf}).

Horizontal long-dashed lines in Fig.~\ref{fig-tdistr}
show $T_\mathrm{s}$ which would have been observed at the same
$T_\mathrm{int}$ if the magnetic field were absent.
In qualitative agreement with the results
of earlier works (Sect.\ \ref{sect-intro}),
the thermal flux emergent from the NS interior
is suppressed by the magnetic fields
in the equatorial region and enhanced near the pole.
For a rotating NS, such distribution leads to a pulsating 
light curve (e.g., Page \cite{page95}),
which may be observed, for instance, with \emph{Chandra}
in soft X-rays (where the spectral flux of the
thermal NS radiation has maximum)
or with the \emph{HST} in UV.

\subsection{Total photon luminosities}
In order to calculate the cooling curves,
one has to know the total NS photon luminosity $L$
at any given $T_\mathrm{int}$. In the non-magnetic case,
the radiation flux emergent from the surface
is given by $F=L/(4 \pi R^2)=\sigma T_\mathrm{s}^4$,
where $T_\mathrm{s} = T_\mathrm{s}^{(0)}(g,T_\mathrm{int})$ 
is known from Eq.~(\ref{PCY-iron}).
In the magnetic case, $F$ should be averaged
over the surface. Using the fitting formulae
(\ref{fit1})--(\ref{fit3}), we have performed such averaging
for the dipole configuration of the field
[Eq.~(\ref{dipole})] at $B_\mathrm{p}\leq10^{16}$~G,
$T_\mathrm{int}=(10^7$--$10^{9.5}$)~K, and $r_g/R\leq0.7$.
The \emph{ratio} of magnetic to non-magnetic fluxes
can be fitted as
\begin{equation}
  {F(B)\over F(0)} = {1 + a_1 \beta^2 + a_2 \beta^3
   + 0.007 \, a_3 \beta^4
       \over
         1 + a_3 \beta^2},
\label{fit-F}
\end{equation}
where
\begin{eqnarray}
    \beta &=&  0.074 \,\sqrt{B_{12}} \,T_\mathrm{int,9}^{-0.45},
 \nonumber\\
    a_1 &=& {5059 \,T_\mathrm{int,9}^{3/4}
            \over
      (1 + 20.4 \,T_\mathrm{int,9}^{1/2}
       + 138 \,T_\mathrm{int,9}^{3/2}
       + 1102 \,T_\mathrm{int,9}^2)^{1/2} },
 \nonumber\\
    a_2 &=& {1484 \, T_\mathrm{int,9}^{3/4}
            \over
      (1 + 90\,T_\mathrm{int,9}^{3/2}
       + 125 \,T_\mathrm{int,9}^2 )^{1/2} },
 \nonumber\\
    a_3 &=& {5530 \,(1 - 0.4 \,r_g/R)^{-1/2}
     \,T_\mathrm{int,9}^{3/4} 
            \over
 (1 + 8.16 \,T_\mathrm{int,9}^{1/2}
 + 107.8 \, T_\mathrm{int,9}^{3/2}
 + 560 \,T_\mathrm{int,9}^2)^{1/2} },
\nonumber
\end{eqnarray}
and $B_{12}$ relates to the magnetic pole.
The maximum error of this fit is 6.1\%\ (i.e., 1.5\%\ for
$T_\mathrm{e}$). Note that 
Eqs.~(\ref{fit1})--(\ref{fit3})
used for calculating $F(B)$
are actually less accurate by themselves, which lowers the
real accuracy of our analytic fits. Nevertheless the 
latter accuracy 
seems to be sufficient for cooling simulations.

Note one important feature: the effect of the
magnetic field on the $F(B)/F(0)$
ratio becomes weaker with growing $T_\mathrm{int}$.
It is explained by the arguments presented
in Sect.\ \ref{sect-profiles}.
Accordingly, the luminosity
of a hot NS cannot be strongly affected even
by very high magnetic fields.

\section{Cooling}
\label{sect-cool}

Let us discuss briefly the effects of
magnetized envelopes on NS cooling.
For illustration, we take the models of
NSs with two masses, $M=1.3$ and $1.5 \, M_\odot$.
For the chosen EOS in the stellar
core (Sect.\ \ref{sect-input}),
the $1.3 \, M_\odot$ NS has the radius $R=11.86$ km.
Its central density $\rho_c=1.070 \times 10^{15}$ g cm$^{-3}$
is lower than the density $\rho_\mathrm{DU}=1.298 \times 10^{15}$
g cm$^{-3}$ at which the powerful direct Urca process
of neutrino emission is switched on. Thus the
main neutrino emission mechanisms in the NS core are
a modified Urca process and neutrino bremsstrahlung in
nucleon-nucleon collisions. Accordingly, the $1.3 \, M_\odot$
model gives us the example of \emph{slow} cooling.
For $M = 1.5 \, M_\odot$ we have $R=11.38$ km
and $\rho_c=1.420 \times 10^{15}$ g cm$^{-3} > \rho_\mathrm{DU}$.
The direct Urca process is open in the central kernel
of mass $0.065 \, M_\odot$ and radius 2.84 km.
This is an example of \emph{fast} cooling.
Further details of these NS models may be found
in GYP.

We have calculated the cooling curves of NSs
with magnetized envelopes, assuming
the dipole magnetic field, Eq.\ (\ref{dipole}).
For comparison with earlier papers (Sect.\ \ref{sect-intro}),
we also consider a familiar model of
the field which is \emph{radial} and constant
in magnitude throughout the blanketing envelope.
In addition, we have switched on and off
superfluidity of neutrons and
protons in stellar interiors to demonstrate the
combined effects of magnetized envelopes and
superfluid interiors on NS cooling.

\subsection{Cooling of non-superfluid neutron stars}
\subsubsection{General features}
\begin{figure}
\centering
\epsfxsize=\hsize
\epsffile{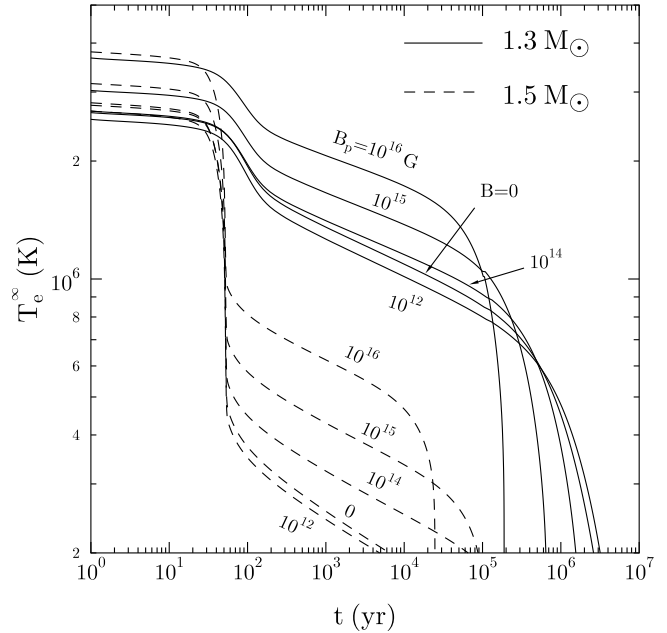}
\caption{Cooling curves of non-superfluid
$1.3 \, M_\odot$ and $1.5 \, M_\odot$ NSs
with dipole magnetic fields
of different strengths.
}
\label{fig-c1}
\end{figure}

Figure \ref{fig-c1} shows the cooling curves
for non-superfluid $1.3\, M_\odot$ and $1.5\, M_\odot$ NSs
with dipole magnetic fields. The curves
are marked with the magnetic field strengths at the magnetic poles,
$B_\mathrm{p}$, and display the effective
temperature $T_\mathrm{e}^\infty$, Eq.\ (\ref{L}).
During the first 50 yr of their lives, $1.3 \, M_\odot$
and $1.5 \, M_\odot$ NSs have nearly the same surface
temperatures since the stellar interiors
are non-isothermal and thermal evolution of the stellar
crust is decoupled from the evolution of the core
(e.g., GYP). Later, 
thermal equilibrium is established in the interior,
and the surface temperature of the $1.5\, M_\odot$ NS becomes much lower
due to very powerful direct Urca processes in the stellar kernel.

The results for $B \la 10^{14}$ G are in good
qualitative agreement with those obtained
by Page (\cite{page95}) and Shibanov \& Yakovlev (\cite{shibyak}).
The thermal state of the stellar interior
is almost independent of the magnetic field in the
NS envelope at the neutrino cooling stage
($t \la 10^4-10^5$ yr), but is affected by
the magnetic field later, at the photon cooling stage.
On the contrary, the surface temperature
is always affected by the magnetic field.
The dipole field $B \la 10^{13}$ G makes
the blanketing envelope overall less heat-transparent.
This lowers $T_\mathrm{e}$ at the neutrino cooling
stage  and slows down cooling at the photon cooling stage.
The dipole field $B \gg 10^{13}$ G makes the
blanketing envelope overall more transparent,
increasing $T_\mathrm{e}$ at the neutrino cooling stage
and accelerating the cooling at the photon stage.

Contrary to the case of the dipole magnetic field,
any radial field (e.g., Van Riper \cite{kvr91})
would always lower the thermal insulation
of the blanketing envelope, increasing $T_\mathrm{s}$
at the neutrino cooling stage and accelerating cooling
at the photon cooling stage. The radial field would affect
the cooling noticeably more than the
dipole field.

Even very strong magnetic fields do not change
significantly thermal insulation of a hot
blanketing envelope (Sect.\ \ref{sect-Tvar}).
Accordingly, the effects
of magnetic fields in a young and hot NSs
are not too strong. The strongest effects
take place in cold NSs, at the photon cooling
stage. Unfortunately, our knowledge of insulating
properties of the blanketing envelopes is the
poorest for these NSs (Sect.\ \ref{sect-input}).
In addition, the magnetic field evolution 
and reheating processes
may become
important for these stars, which we ignore for simplicity.
Notice that the fields $B \ga 10^{14}$ G
appreciably accelerate the cooling and lower
the duration of the neutrino cooling stage.

\subsubsection{Effects of magnetic field and light bending}
\begin{figure}
\centering
\epsfxsize=\hsize
  \epsffile{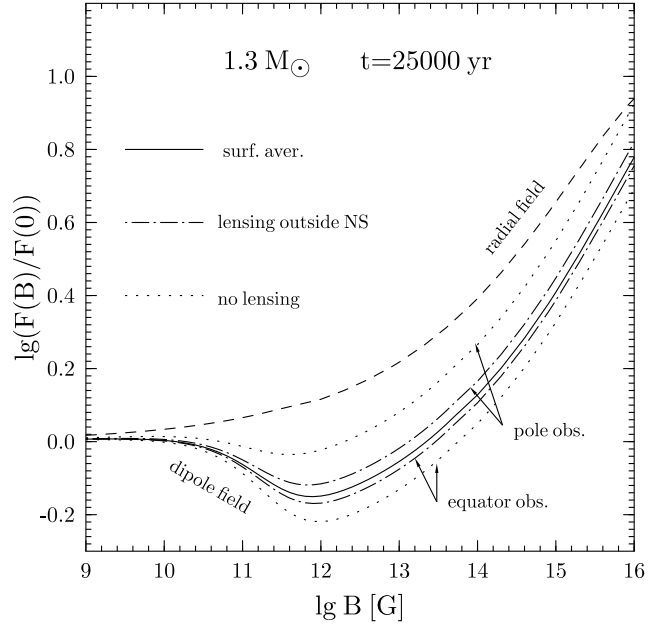}
\caption{Effect of dipole
or radial surface
magnetic fields on the flux of electromagnetic
radiation $F(B)$ detected from a
non-superfluid $1.3 \, M_\odot$
NS of age $t=25000$ yr. Solid line shows the
flux averaged over the entire NS surface
for the dipole field. Dot-and-dashed lines present
the flux observed in the direction towards the
magnetic pole (assumed to be aligned with the
rotational axis) or towards the magnetic equator.
The dotted lines show the same fluxes calculated
neglecting gravitational bending of light rays
outside the NS.
 }
\label{fig-c2}
\end{figure}

Let us consider some effects of the magnetic fields on NS
cooling in more detail. Figure \ref{fig-c2}
demonstrates the effect of dipole or radial
magnetic fields on the flux of electromagnetic radiation
$F(B)$ detected from a non-superfluid $1.3 \, M_\odot$ NS
of an age of 25\,000 yr. This is the age of the Vela pulsar
(Lyne et al.\ \cite{lyneetal96}).
This is also a typical age of anomalous X-ray pulsars
which are likely
to be magnetars (e.g., Mereghetti \cite{mereghetti01}).
Any radial magnetic field is seen to increase the flux.
For the field strength $B = 4 \times 10^{12}$ G,
which can be appropriate for
the Vela pulsar, the increase is by a factor of 1.5,
while for ultrahigh $B =10^{15}$ and
$10^{16}$ G
the factor is about 4.5 and 9, respectively.

For the dipole field, however, the detected flux depends
on observation direction. 
The dipole field
$B_\mathrm{p} =4 \times 10^{12}$ G reduces the 
flux averaged over the entire NS surface (solid line)
by a factor of 1.3, while the fields 
$B_\mathrm{p} =10^{15}$ and $10^{16}$ G
amplify it by factors of
about 2 and 4, respectively.
The dashed
and dotted lines in Fig.~\ref{fig-c2} refer to the fluxes detected
in the direction to the magnetic pole and equator,
respectively (assuming the magnetic axis coincides
with the rotational one, for simplicity). These
lines show the largest difference of the fluxes
detected under different angles.
They are obtained taking proper account of
the general relativity effect of bending of light rays
propagating from the NS surface to a distant observer,
in the same manner as was done by Page (\cite{page95}).
For comparison, we present also the fluxes
calculated neglecting the gravitational ray-bending effect
(as if space-time outside the NS were flat).
In the absence of light
bending effects, the difference of fluxes
observed under different angles is quite
noticeable. For instance, at $B_\mathrm{p}=4 \times 10^{12}$ G
the largest difference would
be about 65\%.
The light bending reduces
the largest difference to 14\%
for $B_\mathrm{p} =4 \times 10^{12}$~G,
making it practically negligible.
This is explained
(e.g., Page \cite{page95}) by the fact that
gravitational bending of light rays increases
the fraction of the NS surface visible by
the observer. The observer detects the flux
from the larger portion of the surface, that
is closer to the flux averaged over observation directions.
The light bending effect is
strong for our $1.3\, M_\odot$ and $1.5\, M_\odot$
NS models. Thus we will neglect weak dependence of
fluxes on the detection direction and use the
average fluxes and associated effective
temperatures $T_\mathrm{e}$ in our analysis.
Notice that here we mean total (spectral integrated)
fluxes but not the fluxes observed in selected
spectral bands taking into account interstellar
absorption.

\begin{figure}
\centering
\epsfxsize=\hsize
 \epsffile{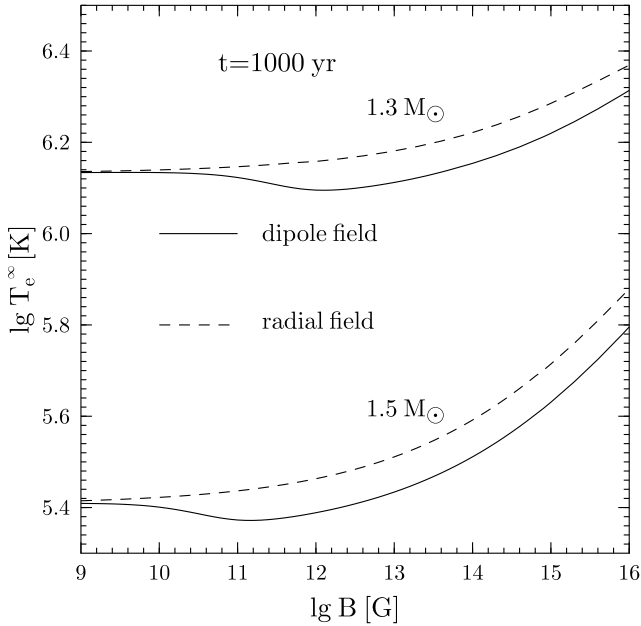}
\caption{Effect of dipole or radial surface
magnetic fields on the averaged
effective surface temperature $T_\mathrm{e}^\infty$
of non-superfluid $1.3 \, M_\odot$ and $1.5 \, M_\odot$
NSs of age $t=1000$ yr.
}
\label{fig-c3}
\end{figure}

Figure \ref{fig-c3} shows the effective
temperatures
of young (1000 yr) non-superfluid $1.3 \, M_\odot$
and $1.5 \, M_\odot$ NSs with dipole and
radial magnetic fields of different strengths.
The effects of magnetic fields are similar to
those in Fig.\ \ref{fig-c2}.
Since the $1.5 \, M_\odot$ NS undergoes fast cooling,
its internal temperature ($\sim 10^7$ K)
is much lower
than in the $1.3 \, M_\odot$ star ($\sim 3 \times 10^8$ K).
Accordingly, the magnetic fields
affect the surface temperature of the $1.5 \, M_\odot$
NS more significantly. Previously,
it was suggested (e.g., Heyl \& Hernquist \cite{HH97})
that the models of
young and hot, slowly
cooling NSs with ultramagnetized envelopes 
could serve as models of anomalous X-ray pulsars and
soft gamma repeaters.
Although those previous cooling models were rather simplified, 
they are in qualitative agreement with our
more elaborate models.
The main outcome of these studies is that even
ultrahigh magnetic fields cannot change appreciably
the average surface temperatures of young
NSs with iron envelopes (as seen clearly 
from Figs.\ \ref{fig-c2} and \ref{fig-c3}).
We stress that the
\emph{local} surface temperatures in narrow strips
near the magnetic equators appear to be much
lower than near the magnetic poles (Sect.\ \ref{sect-thst}),
but they do not contribute noticeably to the
NS luminosity integrated over the surface.

\subsection{Effects of superfluidity 
and application to the Vela pulsar}
\begin{figure}
\centering
\epsfxsize=\hsize
\epsffile{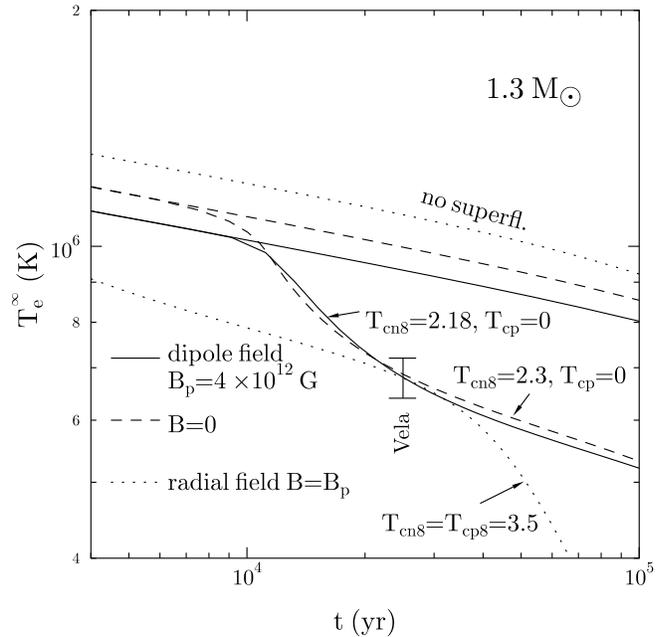}
\caption{Cooling of $1.3 \, M_\odot$ NS
   with $B=0$, with the dipole
   magnetic field  $B_\mathrm{p}= 4 \times 10^{12}$ G
   or with the radial field of the same strength.
   The upper curves are calculated assuming
   no superfluidity in the NS interior
   while the lower ones are calculated
   for certain critical temperatures of the neutron and
   proton superfluids in the NS cores
   (the values of $T_\mathrm{cn}$ and $T_\mathrm{cp}$
   are indicated near the curves in units of $10^8$~K).
   Error bars show the possible interval of the surface
   temperatures of the Vela pulsar (see text).
}
\label{fig-c4}
\end{figure}

%
%
Our analysis shows that
the magnetic fields of ordinary pulsars
($B \la 10^{13}$ G) do not strongly affect the
surface temperature, at least at the neutrino
cooling stage. Nevertheless, such fields may influence 
the theoretical
interpretation of observations. For illustration,
let us analyse observations of the thermal radiation
from the Vela pulsar. We adopt the value of
the effective surface temperature
$T_\mathrm{e}^\infty=  (6.8 \pm 0.4) \times 10^5$ K
inferred most recently 
from observations with the \emph{Chandra} observatory 
by Pavlov et al.\ (\cite{Vela-GG})
(at the 1$\sigma$ level)
using the hydrogen atmosphere models.
This value of $T_\mathrm{e}^\infty$
is in a good agreement with the
value $T_\mathrm{e}^\infty=  (7.85 \pm 0.25) \times 10^5$ K
obtained earlier by Page et al.\ (\cite{pageetal96})
from \emph{ROSAT} observations. 
Thus we assume the presence of a very thin
hydrogen atmosphere (of mass $\la 10^{-13} \, M_\odot$)
which determines the spectrum of thermal radiation
from the NS surface but does not affect thermal
insulation of a blanketing envelope composed
mostly of iron. Let us use the slowly cooling
$1.3 \, M_\odot$ NS model for the interpretation
of observations. Assuming 
$B=0$ and no superfluidity in the NS interior,
we have, for the age of 25\,000 yr,%
\footnote{We have verified that $T_\mathrm{e}^\infty$
is almost independent of the NS mass, for the given EOS, 
as long as $M$ is lower than
the critical value $1.44 \, M_\odot$ at which the direct
Urca process becomes open in the stellar interior.}
$T_\mathrm{e}^\infty\approx1.0\times10^6$ K,
which is higher than the value
obtained from observations (Fig. \ref{fig-c4}).
The dipole magnetic field $B_\mathrm{p} = 4 \times 10^{12}$ G
lowers $T_\mathrm{e}^\infty$
by 8\%\ (Fig.\ \ref{fig-c2}), not sufficient to attain
the required value.

It is well known that
neutron or proton superfluidity in the NS core
affects the cooling.
For illustration, we adopt the same simplified model
of nucleon superfluidity which has been used in
a number of previous works
(Yakovlev et al.\ \cite{yaketal99} and references therein).
In this model, the neutron pairing is
assumed to be in the triplet state, while the
proton pairing occurs in the singlet state of
a nucleon-nucleon pair. 
The critical temperatures $T_\mathrm{cn}$
and $T_\mathrm{cp}$ of the neutron and proton superfluids are
assumed to be constant and treated as free parameters
to be adjusted to observational data.
Figure \ref{fig-c4} demonstrates that taking
the neutron and proton superfluidities with certain
values of $T_\mathrm{cn}$ and $T_\mathrm{cp}$
we can lower $T_\mathrm{e}$ of our
Vela model to the observed values. This can be done for all
three models of the envelope, with $B=0$ as well as with the
dipole and radial fields. In the given examples,
the lowering of $T_\mathrm{e}$ 
for the dipole or field-free cases is 
produced by neutrino emission due to Cooper
pairing of neutrons, while for the radial
field the lowering is produced by neutrino
emission due to Cooper pairing of neutrons and protons.
In these three cases
we need different critical temperatures $T_\mathrm{cn}$
and $T_\mathrm{cp}$. In each case the choice of $T_\mathrm{cn}$
and $T_\mathrm{cp}$ is not unique.

\begin{figure}
\centering
\epsfxsize=\hsize
\epsffile[35 175 470 590]{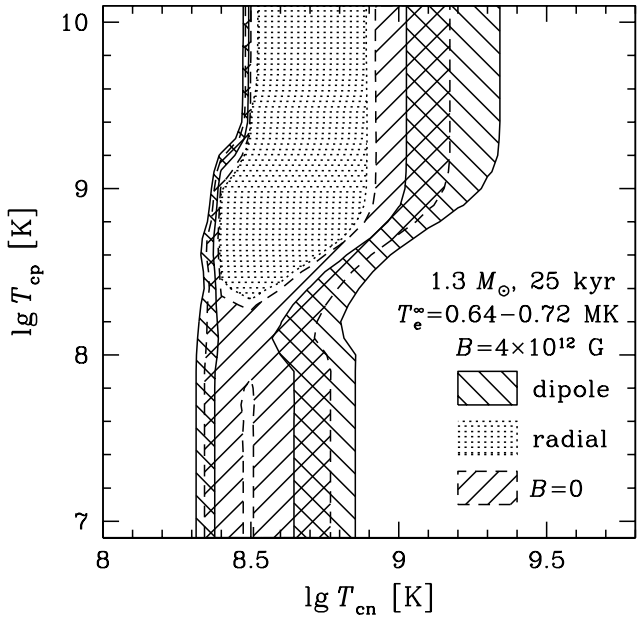}
\caption{Domains (hatched)
  of $T_\mathrm{cn}$ and $T_\mathrm{cp}$
  for which the $1.3 \, M_\odot$ NS
  with the superfluid core has
  the surface temperature
  $T_\mathrm{e}^\infty=(6.8 \pm 0.4) \times 10^5$ K
  by the age of the Vela pulsar.
  Solid lines enclose the domains
  for the NS with the dipole magnetic field ($B_p=4 \times 10^{12}$ G);
  dashed lines are for the non-magnetized NS
  and dots are for the radial field ($B=B_p$).
}
\label{fig-c5}
\end{figure}

%
Generally, if we fix the blanketing envelope model,
we can determine the domains of $T_\mathrm{cn}$ and
$T_\mathrm{cp}$ values in the $T_\mathrm{cn}$--$T_\mathrm{cp}$ plane
which force the NS
to have $T_\mathrm{e}^\infty$ within the
given errorbar by the specified age $t$.
These domains (hatched regions) are shown in Fig.\ \ref{fig-c5}
for the Vela pulsar with a non-magnetic envelope,
and also for the envelope
with dipole and radial magnetic fields.

The domain for the non-magnetic envelope 
has a complicated X-shape. 
The dipole magnetic field
facilitates a lowering of $T_\mathrm{e}$ to the
required errorbar.
The domain of acceptable values
of $T_\mathrm{cn}$ and $T_\mathrm{cp}$ becomes narrower overall
and splits into two parts (Fig.\ \ref{fig-c5}).
In principle, in these two cases we do not need
proton superfluidity to explain the
observations (Fig.\ \ref{fig-c4}): the neutrino
emission due to Cooper pairing of neutrons is
sufficiently strong by itself to lower $T_\mathrm{e}$.
On the other hand, were the magnetic field radial, it would
raise $T_\mathrm{e}$ and complicate
the adjusting of $T_\mathrm{e}$ to observations.
Neutrino emission due to Cooper pairing of neutrons is
insufficient to lower $T_\mathrm{e}$, and the 
emission due to Cooper pairing of protons
is needed. The domain acquires a qualitatively
different shape: the proton superfluidity
cannot be too weak ($\lg T_\mathrm{cp} \ga 8.3$).

The domains of $T_\mathrm{cn}$ and $T_\mathrm{cp}$ obtained
for the Vela pulsar
can be combined further with analogous domains
determined for other cooling NSs in attempt
to impose restrictions on the values of
$T_\mathrm{cn}$ and $T_\mathrm{cp}$ in the stellar cores.
A preliminary analysis of such a kind has been done 
by Yakovlev et al.\ (\cite{yaketal99}). 
However, further work remains to be done
to constrain strongly the critical temperatures:
one should take into account new observational data
on thermal radiation from isolated NSs and
new theoretical models of cooling NSs
with superfluid cores; particularly,
one has to incorporate density dependence of
$T_\mathrm{cn}$ and $T_\mathrm{cp}$ over the stellar core.
Such work is beyond the scope of the present
paper, but our illustrative examples show that
proper analysis of the NS superfluidity may require
consideration of
the effects of the magnetic field on the NS blanketing envelope.

\section{Conclusions}

We have analysed the thermal structure of neutron star
envelopes with typical pulsar magnetic fields
$10^{11}$--$10^{13}$ G and with ultrahigh
magnetic fields up to $10^{16}$ G.
We have used (Sect.\ \ref{sect-input})
modern data on equation of state
and thermal conductivities of magnetized neutron star
envelopes. In particular, we have proposed
an analytic model of the radiative thermal
conductivity limited by the Thomson scattering
and free-free absorption of photons in a magnetized
plasma. We have used the values of thermal conductivity
of degenerate electrons, updated recently by
proper treatment of dynamical ion-ion
correlations which affect electron-ion scattering.
We have calculated the temperature profiles
(Sect.\ \ref{sect-thst})
in the neutron star envelope for any
inclination of the magnetic field to the surface
and obtained a fit expression
which relates the internal temperature of the neutron
star to the local effective surface temperature.
We have also calculated and fitted the
relation between the internal temperature and
total surface luminosity (or effective
temperature) for the dipole magnetic fields
$B \la 10^{16}$ G. 

Furthermore, we have performed (Sect.\ \ref{sect-cool}) 
simulations of cooling
of neutron stars with dipole or radial magnetic fields in the envelopes.
In agreement with the previous studies of
Page (\cite{page95}) and Shibanov \& Yakovlev (\cite{shibyak}),
we have found that the effects 
of the two magnetic field configurations
on neutron star cooling
are qualitatively different. We have briefly discussed
cooling of young and middle-aged neutron stars
with ultramagnetized envelopes (magnetars as the models
of soft gamma repeaters and anomalous
X-ray pulsars) and the effect of surface
magnetic fields on constraining critical
temperatures of the neutron and proton superfluids
in the cores of ordinary pulsars, considering the
Vela pulsar as an example.

We stress that our results are less
reliable for cold neutron stars
($T_\mathrm{s} \la 3 \times 10^5$ K) with
very strong magnetic fields $B \ga 10^{13}$ G
because of lack of knowledge of ionization
state and thermal conductivity of
the outermost parts of the cold blanketing envelopes.
Further work is required to fill these gaps. 
In this paper, we have considered
the blanketing envelope composed of
iron and have not analysed the envelopes
containing light elements.
The presence of light elements generally
reduces thermal insulation of the envelope.
This effect has been studied in detail
for non-magnetic envelopes (PCY)
and is expected to be important
for magnetic envelopes as well (Heyl \&
Hernquist  \cite{HH97}). We are planning to address
this matter in a future work.

\begin{acknowledgements}
We are grateful to G.\ Pavlov for encouragement and for
providing us with the
results of observations (Pavlov et al.\ \cite{Vela-GG})
before publication, to U.\ Geppert 
for stimulating suggestions, 
to Yu.A.\ Shibanov and K.P.\ Levenfish for
helpful discussions, and to the referee, Dany Page,
for useful remarks.
This work was supported by RFBR (grant No.\ 99-02-18099).
\end{acknowledgements}

\end{document}